\documentclass[a4paper,12pt]{article}

\usepackage[latin1]{inputenc}
\usepackage{amsmath}
\usepackage{amsfonts}
\usepackage{amssymb}
\begin{document}

\newcommand{\tr}{\mathrm{tr}}

\begin{titlepage}

\topmargin=3.5cm

\textwidth=13.5cm

\centerline{\Large \bf Equivariant Localization}

\vspace{0.4cm}

\centerline{\Large \bf  in Stochastic Quantization and}

\vspace{0.4cm}
\centerline{\Large \bf  Quenched Matrix Models}

\vspace{1.0cm}

\centerline{\large \bf Levent Akant\footnote{E-mail:
akant@gursey.gov.tr}}

\vspace{0.5cm}

\centerline{ \textit{Feza Gursey Institute}}

\centerline{\textit{Emek Mahallesi, Rasathane Yolu No.68 }}
\centerline{\textit{Cengelkoy, Istanbul, Turkey}}

\vspace{1.5cm}

\paragraph{Abstract:} It is shown that Parisi-Sourlas supersymmetry of stochastic quantization is a Cartan model of equivariant cohomology. Equivariant cohomological structure of stochastic quantization of linear and non-linear sigma models are discussed.
Witten's nonabelian localization principle is applied to the stochastic quantization of matrix models. As a result the equivalence between the original matrix model and the corresponding quenched Eguchi-Kawai model is established.

\end{titlepage}

\section{Introduction}

Our aims in this paper are (i) to interpret the Parisi-Sourlas
\cite{p2} supersymmetry of stochastic quantization \cite{pw} as a
Cartan model of equivariant cohomology and (ii) to derive quenched
Eguchi-Kawai matrix models \cite{ek, bhn} by applying equivariant
localization principle to the stochastic quantization of matrix
models. The equivariant localization principle that we will use in
the present work is Witten's nonabelian localization principle
\cite{witten1, KJ, witten2}. Although the group which gives rise to the Cartan
model that we will use is abelian, the methods of nonabelian
localization are more natural for the questions we will address
here. The methods we will employ here will be similar to those developed in \cite{akant} to interpret gauge fixing procedure as equivariant localization. In particular we will not use any supergeometry \cite{sz}. The use of
localization principle in dimensional reduction was, to the best of
our knowledge, first proposed by Zaboronsky in \cite{zabo}.

The basic idea behind stochastic quantization is to compute the Green functions of a $d$ dimensional Euclidean field theory as the equilibrium expectation values of a statistical system evolving with stochastic time $t$. The approach to equilibrium in the limit $t \rightarrow \infty$ is assumed to be governed by a Langevin equation with an external Gaussian random source (white noise). The interesting point is that even before we take the large $t$ limit, the correlations of the Langevin equation can be calculated as Green's functions of a local field theory living in $d+1$ dimensions. The action of this theory, which is often called the Fokker-Planck action, is invariant under certain supersymmetry transformations known as the Parisi-Sourlas supersymmetry. We shall show that this supersymmetry can be interpreted as a Cartan model of equivariant cohomology. This will allow us to view the Fokker-Planck action as a closed equivariant form and to apply the equivariant localization principle to localize the correlations of the theory.

Given a matrix model in $d$ dimensions, quenching prescription relates \cite{ek, bhn}, in the large N limit, the correlations of the original theory to the correlations of a $0$ dimensional matrix model. We will show that the quenching prescription is a consequence of the equivariant localization principle as applied to the Fokker-Planck action. Thus we will provide an example of a situation in which the equivariant localization principle is used to establish the equivalence of two distinct field theories living in different dimensions.

Matrix models are field theories whose degrees of freedom can be
represented by (space-time dependent) matrices. Such an arrangement
of degrees of freedom has the merit of making certain internal
symmetries of the model manifest. In this paper we will consider
Hermitian matrix models where the degrees of freedom of the model
are arranged into Hermitian matrices. Moreover we will consider actions
which are invariant under conjugation by the unitary group. Thus a
typical action will be of the form
\begin{equation}
S=\int d^{d}x \tr\left[ \frac{1}{2}\partial_{\mu}\phi\partial^{\mu}\phi+V(\phi)\right].
\end{equation}
Here $\phi(x)$ is the $N\times N$ Hermitian matrix field and
$V(\phi)$ is a polynomial in $\phi$. Clearly the action is invariant
under the action of $U(N)$:
\begin{equation}
\phi(x)\rightarrow U^{\dagger}\phi(x)U.
\end{equation}
We will often refer to this symmetry as the color symmetry.

It is well known that matrix models simplify in the large $N$ limit. This simplification is mainly due to the fact that in the large $N$ limit the expectation value of a product of color invariant observables factorizes into a product of expectation values. This observation allows us to treat the large $N$ limit as a classical limit. Diagrammatically the large $N$ limit corresponds to the summation of all planar graphs. Since a planar diagram may contain an arbitrarily large number of loops, we see that this classical limit is not the usual $\hbar\rightarrow 0$ limit.

Another important property of the large $N$ limit is that it relates
the correlations of color invariant observables of a $d$ dimensional
matrix model to the expectation values of the so-called quenched
observables of a certain $0$ dimensional reduced matrix model. The
reduced models in question were first derived by Eguchi and Kawai in
\cite{ek}. Whereas the correct observables to use in the reduced
model and the subsequent quenching prescription were discovered by
Bhanot, Heller and Neuberger in \cite{bhn} and further studied in
\cite{parisi1, dw, gk, as, gh, das}. Explicitly the quenching
prescription goes as follows. Consider a Hermitian matrix model. Let
$F$ be a color invariant observable of the form
\begin{equation}
F=\frac{1}{N}\tr\left[\phi(x_{1})\ldots \phi(x_{n}) \right].
\end{equation}
Then in the large $N$ limit
\begin{eqnarray}\label{quench}
<F>=\int\left(\prod_{\mu,a}dp^{a}_{\mu} \right)\rho(P)<\widetilde{F}>_{EK}
\end{eqnarray}
Here
\begin{equation}
\widetilde{F}=\tr\left[\widetilde{\phi}(x_{1})\ldots \widetilde{\phi}(x_{n}) \right].
\end{equation}
\begin{eqnarray}
\widetilde{\phi}(x)=e^{iP\cdot x}Ae^{-iP\cdot x},\;\,\;\; (P^{\mu})^{a}_{b}=p^{\mu}_{a}\delta^{a}_{b},
\end{eqnarray}
with space-time independent $A$ and $P$. The combinations
$\widetilde{\phi}(x)$ are called quenched fields. $\rho(P)$ is a
normalized Gaussian. The expectation value $<\;\;>_{EK}$ is with
respect to the $0$ dimensional reduced model with the action
\begin{equation}
S_{EK}[A]=\tr\left(-\frac{1}{2}[P_{\mu},A][P_{\mu},A]+V(A) \right).
\end{equation}
The real degree of freedom of this action is the random matrix $A$.
However, one also integrates over $P$. But this integration is only
after the calculation of the expectation value with respect to
$S_{EK}$. This procedure is similar to the quenching prescription of
statistical mechanics, hence the name quenched matrix models.

So we see that quenching prescription relates the correlations of
two theories living in different dimensions. There are various
arguments for the validity of the quenching prescription which can
be found in the references given above. We will not repeat those
derivations here. Instead we will give an alternative geometric
point of view based on the use of equivariant localization principle
in the stochastic quantization of matrix models.

\section{Stochastic Quantization}

The basic idea behind stochastic quantization (for a collection of reprints see \cite{stochastic}) is to compute the Green functions of an Euclidean field theory as the equilibrium expectation values of a statistical system evolving with stochastic time $t$. This evolution is assumed to be governed by a
Langevin equation with an external Gaussian random source. More precisely let the action for the system in equilibrium be denoted by $S[\phi]$, then the equilibrium expectation values of the random fields $\phi(x)$ are given by the large $t$ limit of fields $\phi(t, x)$ evolving according to the Langevin equation
\begin{equation}
\frac{\partial\phi(t,x)}{\partial t}+\frac{\delta
S[\phi]}{\delta\phi(t,x)}=\eta(t,x).
\end{equation}
This is a stochastic differential equation where the source $\eta(t,x)$ is a random variable with a Gaussian distribution
\begin{equation}
\mathcal{D}\eta(t,x)e^{-\int dt d^{d}x \;\eta^{2}(t,x)}.
\end{equation}
The correlation functions of the Langevin equation are defined as
\begin{equation}
<\phi_{\eta}(t_{1},x_{1})...\phi_{\eta}(t_{n},x_{n})>_{\eta}=\int \mathcal{D}\eta\,e^{-\int dt dx \eta^{2}(t,x)}
\phi_{\eta}(t,x_{1})...\phi_{\eta}(t,x_{n})
\end{equation}
where $\phi_{\eta}$ is the perturbative solution in $\eta$ of the Langevin equation. It is a random variable since the source $\eta$ is itself random. The important observation made by Parisi and Wu in \cite{pw} was that
in the $t \rightarrow\infty$ limit these correlation functions approach the Green's functions of our field theory with action $S[\phi]$
\begin{equation}
\lim_{t\rightarrow \infty}<\phi_{\eta}(t, x_{1})...\phi_{\eta}(t, x_{n})>_{\eta}=\frac{1}{Z}\int\mathcal{D}\phi\,\phi(x_{1})...\phi(x_{n})\;e^{-S[\phi]}.
\end{equation}
Here $Z$ is the partition function of $S[\phi]$. This can be proven perturbatively or by making use of the Fokker-Planck equation associated with the Langevin equation. In the case of gauge theories the identity holds only for the gauge invariant observables.

A closer analysis of the correlations of the Langevin equation shows that they are the Green's functions of a field theory in $d+1$ dimensions with the action
\begin{eqnarray}
S_{FP}&=&\int dt d^{d}x \, \frac{1}{2}b^{2}(t,x)-ib(t,x)\eta_{\phi}\nonumber\\
&&+\overline{\psi}(t',x')\left[
\partial_{t}\delta(t-t')\delta^{d}(x-x')-\frac{\delta^{2}S[\phi]}{\delta\phi(t',x')\delta\phi(t,x)}\right]
\psi (t,x)
\end{eqnarray}
where
\begin{equation}
\eta_{\phi}=\frac{\partial\phi(t,x)}{\partial t}+\frac{\delta
S[\phi]}{\delta\phi(t,x)}
\end{equation}
We will give a derivation of this action in Appendix A. This action is usually called the Fokker-Planck (FP) action.
It is well known that FP action is invariant under the Parisi-Sourlas supersymmetry \cite{p2}
\begin{eqnarray}
Q\phi&=&\psi\\
Q\psi&=&0\\
Q\overline{\psi}&=&-ib\\
Qb&=&0.
\end{eqnarray}

In the case of a non-linear sigma model where the fields $\phi^{i}$
obey nonlinear constraints the Langevin equation should be modified.
The correct modification and the resulting FP action were given by
Zinn-Justin in \cite{zj}. Here we will only quote the resulting FP
action:
\begin{equation}
S_{FP}=\overline{\rho}+\overline{\omega}
\end{equation}
where
\begin{eqnarray}
\overline{\rho}&=&\int d^{d}x dt \,\left[ \frac{1}{2}\partial_{i}\partial_{j}g_{kl}(\phi)\overline{\psi}^{i}\psi^{j}\overline{\psi}^{k}\psi^{l}-
ib^{i}g_{in}(\phi)\Gamma^{n}_{kl}(\phi)\overline{\psi}^{k}\psi^{l}+\frac{1}{2}g_{ij}b^{i}b^{j}\right]\nonumber\\
\overline{\omega}&=&-\int d^{d}x dt\, \overline{\psi}^{i}\left[g_{ij}(\phi)\frac{\partial}{\partial t}+\partial_{j}g_{ik}(\phi)
\frac{\partial\phi^{k}}{\partial t}+\frac{\delta^{2} S}{\delta\phi^{i}(t,x)\delta\phi^{j}(t,x)}\right]\psi^{j}\nonumber\\
&&-i\int d^{d}x dt\, b^{i}\mu_{i}\nonumber\\
\mu_{i}&=&g_{ij}\frac{\partial \phi^{j}}{\partial t}+\frac{\delta
S}{\delta \phi^{i}(t,x)}.
\end{eqnarray}
Here $g_{ij}$, $\Gamma^{i}_{jk}$ and $R_{ijkl}$ are respectively the metric, the Christoffel symbols and the Riemannian curvature of the target manifold. This action is still invariant under the BRST transformations given above.

\section{Equivariant Cohomology}
In this section we will give a brief summary of equivariant
cohomology and equivariant localization principle \cite{witten1, DH,
atiyah-bott}. Equivariant cohomology can be thought of as
an extension of the de Rham cohomology when there is a group action
on the manifold underlying the de Rham complex. If the group action
is free then the equivariant cohomology is the same as the de Rham
cohomology of the quotient manifold. On the other hand if the action
is not free the quotient space is not a smooth manifold. In this
case a sensible cohomology theory can be constructed using
classifying bundles. This type of construction often involves
infinite dimensional manifolds which are hard to deal with. However
there exist differential complexes with finitely many generators
whose cohomologies are equivalent to the de Rham cohomology obtained
by using classifying bundles. Here we will work with one such model,
namely the Cartan model of equivariant cohomology. Detailed
reviews of the equivariant cohomology theory can be found in \cite{bott, GS, libine,
moore}.

\subsection{Cartan Model of Equivariant Cohomology}

Let $M$ be a manifold and $G$ be a group acting on $M$. Let us denote the Lie algebra of $G$ by $\mathbf{g}$ and its dual by $\mathbf{g}^{\ast}$ . We will let the dimensions of $M$ and $G$ be $m$ and $n$, respectively.
This data is sufficient to define the Cartan model of equivariant cohomology whose underlying graded differential complex is
\begin{equation}
\mathcal{S}(\mathbf{g^{\ast}})\otimes \Omega(M)
\end{equation}
Here $\mathcal{S}(\mathbf{g^{\ast}})$ is the symmetric tensor algebra on $\mathbf{g}^{\ast}$. A generic element of this tensor product may be represented as a linear combination of exterior forms whose coefficients are polynomials in $n$ variables $b^{a}$, ($a=1,2,\ldots n$). So a generic element is a linear combination over $\mathbf{R}$ of elements of the form
\begin{equation}
b^{I}\alpha^{p}.
\end{equation}
Here $I$ is a multi-index and $\alpha^{p}$ is an exterior $p$ form
on $M$. The grading on $\mathcal{S}(\mathbf{g^{\ast}})\otimes
\Omega(M)$ is defined by assigning degree $2\vert I\vert+p$ to
$b^{I}\alpha^{p}$. Here $\vert I\vert$ is the length of $I$.

The action of $\mathbf{g}$ on $\mathcal{S}(\mathbf{g^{\ast}})\otimes \Omega(M)$ is given by
\begin{equation}
\delta_{\xi}(b^{I}\alpha^{p})=\delta_{\xi}(b^{I})\alpha^{p}+b^{I}\pounds_{V_{\xi}}\alpha^{p}
\end{equation}
where
\begin{equation}
\delta_{\xi}(b^{i_{1}}\ldots b^{i_{r}})=-\sum_{k=1}^{r}b^{i_{1}}\ldots (ad_{\xi}^{\ast}b^{i_{k}})\ldots b^{i_{r}}.
\end{equation}
Here $ad^{\ast}$ is the coadjoint action of $\mathbf{g}$ on $\mathbf{g^{*}}$, $V_{\xi}$ is the fundamental vector field on $M$ corresponding to $\xi \in \mathbf{g}$, and $\pounds$ denotes the Lie derivative. If $\left\lbrace e_{a}\right\rbrace $ is a basis for $\mathbf{g}$ then we will denote $V_{e_{a}}$ simply by $V_{a}$. Moreover we will write $\pounds_{a}=\pounds_{V_{a}}$ and
$\iota_{a}=\iota_{e_{a}}$. Here $\iota$ is the contraction operator on $M$.

The Cartan derivative on $\mathcal{S}(\mathbf{g^{\ast}})\otimes \Omega(M)$ is defined as
\begin{equation}
D=d-ib^{a}\iota_{a}
\end{equation}
One can show that $D^{2}=-i\delta_{v}$. Thus if we consider
$(\mathcal{S}(\mathbf{g^{\ast}})\otimes \Omega(M))^{G}$, the
subspace of $\mathcal{S}(\mathbf{g^{\ast}})\otimes \Omega(M)$ which
is annihilated by all $\delta_{\xi}$, then on it $D^{2}=0$. Thus we
can define a cohomology theory on
$(\mathcal{S}(\mathbf{g^{\ast}})\otimes \Omega(M))^{G}$ with the
Cartan derivative as the differential. This cohomology is called the
Cartan model of equivariant cohomology.

Interesting examples of this construction arise when one takes $M$ to be a symplectic manifold and assumes the action of $G$ to be Hamiltonian. In this case one can give explicit examples of equivariant forms. An important example of a closed equivariant 2-form is the equivariant symplectic form
\begin{equation}
\overline{\omega}=\omega-ib^{a}\mu_{a}.
\end{equation}
It is easy to show that this form is $G$ invariant and satisfies $D\overline{\omega}=0$. Conversely any equivariant 2-form must be of the form $\nu-ib^{a}f_{a}$. Here $\nu$ is a differential 2-form on $M$ and $f$'s are in $C^{\infty}(M)$. Assuming $\nu$ is non-degenerate one can show that the $G$ invariance and closedness of this equivariant 2-form imply that $\nu$ is a symplectic form  and that $G$ has a Hamiltonian action on the symplectic manifold $(M,\nu)$.

Another example of a closed equivariant form is the 4-form constructed using the Cartan-Killing metric on $\mathbf{g}$
\begin{equation}
\frac{1}{2}(b,b)
\end{equation}
From now on we will assume that $\mathbf{g}$ is semi-simple and compact. Thus one can choose the Cartan-Killing metric as $\delta_{ab}$ and the structure constants $c^{abc}=\delta^{dc}c^{ab}_{d}$, completely antisymmetric.

\subsection{Integration and Localization}

An equivariant form
\begin{equation}
a=\sum_{k}P_{k}(b)\alpha^{k}
\end{equation}
can be integrated as
\begin{eqnarray}
\int a=\sum_{k}\int \left(\prod_{a}db^{a} \right)e^{-\frac{1}{2}\epsilon(b,b)}P_{k}(b)\int_{M} \alpha^{k}.
\end{eqnarray}
Here we are assuming $M$ to be orientable. Then the second integral on the right hand side is the usual integration of differential forms on $M$.
In particular it can be nonzero only for the top form $\alpha^{m}$ ($m=\mathrm{dim}M$). Moreover notice that the exponential of the equivariant 4-form $(b,b)$ acts as a regulator for integration over $b$.

A very important property of this integral is that on a compact manifold without boundary the integral of an exact equivariant form vanishes \cite{witten1}. If the manifold is not compact the same result holds if the equivariant form has compact support or decreases fast enough outside a bounded region. A simple but very important consequence of this result is that for any equivariant 1-form $\lambda$ and closed equivariant form $a$
\begin{equation}
\int a=\int ae^{tD\lambda}.
\end{equation}
This is true because $a(1-e^{tD\lambda})$ is an exact equivariant form. This last formula is the basis of the equivariant localization principle. Since the integral on the left hand side is independent of $t$ its value can be calculated as the large $t$ limit of the right hand side
\begin{equation}
\int a=\lim_{t\rightarrow \infty}\int ae^{tD\lambda}.
\end{equation}
However $D\lambda=d\lambda-ib^{a}\lambda(V_{a})$. So in the large $t$ limit the integral localizes
on the critical points of $b^{a}\lambda(V_{a})$
\begin{eqnarray}
\lambda(V_{a})&=&0\\
b^{a}d\lambda(V_{a})&=&0.
\end{eqnarray}
Notice that we are free to choose $\lambda$. One can use this
freedom to simplify things \cite{witten1}. Let $J$ be an almost
complex structure compatible with $\omega$ i.e.
$\omega(JX,JY)=\omega(X,Y)$ and $g(X,Y):= \omega(JX,Y)$ is a
positive definite metric. Let $I=(\mu,\mu)$ and choose
\begin{equation}
\lambda=JdI.
\end{equation}
Then one can show that \cite{witten1}
\begin{equation}\label{le1}
\lambda(V_{a})=0\;\;\Leftrightarrow\;\; dI=0
\end{equation}
which admits two types of solutions i) ordinary critical points: $\mu=0$ or ii) higher critical points: $dI=0$ but $\mu\neq 0$.
One can also show that for ordinary critical points
\begin{equation}\label{le2}
b^{a}\left. d\lambda(V_{a})\right \vert_{\mu^{-1}(0)}=0\;\;\Leftrightarrow\;\;b^{a}=0.
\end{equation}
Let us note that the non-degeneracy of the symplectic form and the positive definiteness of $g$ are essential for the validity of these results \cite{witten1}. Notice also that because of (\ref{le2}) the integral over $b$ localizes on a neighborhood
of $b=0$ where there is no need for the regulator. In fact we will see that in our applications this regulator may be omitted without changing the value of the integral. 
\section{Examples of Closed Equivariant Forms in Local Riemannian Geometry}

In this section we want to give examples of closed equivariant forms
which have their origins in the local geometry of a Riemannian
manifold. Let $U$ be a coordinate chart in a Riemannian manifold $N$
and $V$ be a vector space with $\dim V=\dim N=n$. Let us consider
the even dimensional product manifold $M=U\times V$. $V$ acts on
this product manifold by fiber translations along $V$. If we denote
the coordinates in $U$ by $x^{i}$ and the coordinates in $V$ by
$p^{i}$ then a typical fiber translation transforms these as
\begin{eqnarray}
x^{i}&\rightarrow& x^{i}\\
p^{j}&\rightarrow& p^{j}+a^{j}.
\end{eqnarray}
Clearly the differentials $\psi^{i}:=dx^{i}$ and $\overline{\psi}^{i}:=dp^{i}$ remain invariant under fiber translations. We can use this simple observation to produce equivariant forms on $M$.

\subsection{An Equivariant 2-form in Local Riemannian Geometry}
Our first example is constructed as follows. Let $\omega_{ij}$ be an $n\times n$ matrix valued function on $U$ satisfying
\begin{equation}\label{a1}
\partial_{k}\omega_{ij}+\partial_{j}\omega_{ik}=0.
\end{equation}
Here $\partial_{i}=\frac{\partial}{\partial x^{i}}$. Now on $U\times V$ define the following 2-form (we will often omit the $\wedge$ sign)
\begin{equation}
\omega=\omega_{ij}\overline{\psi^{i}}\wedge \psi^{j}.
\end{equation}
Because of (\ref{a1}) this is a closed form. It is also invariant under the action of $V$. Its equivariant extension $\overline{\omega}=\omega-ib^{a}\mu_{a}$ can be determined by solving
\begin{eqnarray}
\pounds_{i}\mu_{j}&=&0\\
d\mu_{i}&=&-\iota_{i}\omega.
\end{eqnarray}
Here we use the notation
$\pounds_{i}=\pounds_{\frac{\partial}{\partial p^{i}}}$ and
$\iota_{i}=\iota_{{\frac{\partial}{\partial p^{i}}}}$. The first
equation implies that $\mu_{i}$ is independent of $p$'s. Then the
second equation can be integrated to yield $\mu_{i}$. Notice that we
allow the possibility that $\omega_{ij}$ is not invertible at
certain points of $U$. Thus $\omega$ may not be a symplectic form on
$U\times V$. We will have more to say about this point in the next
section. Notice that
\begin{eqnarray}
\iota_{i}\omega=-d\mu_{i}\Rightarrow \omega_{ij}\psi^{j}=-\partial_{j}\mu_{i}\psi^{j}
\end{eqnarray}
Therefore
\begin{eqnarray}
\omega&=&-\partial_{j}\mu_{i}\overline{\psi}^{i}\psi^{j}\nonumber\\
&=&\partial_{j}\mu_{i}\psi^{j}\overline{\psi}^{i}\nonumber\\
&=&d\left\lbrace \mu_{i}\overline{\psi}^{i}\right\rbrace.
\end{eqnarray}
Moreover,
\begin{equation}
\overline{\omega}=D\left\lbrace \mu_{i}\overline{\psi}^{i}\right\rbrace.
\end{equation}
For the moment let us assume that $\omega$ is a symplectic form and define the following almost complex structure which is invariant under the action of $V$
\begin{eqnarray}
J=\sum_{ij}\,2\omega^{ij}\psi^{j}\otimes \frac{\partial}{\partial p^{i}}-\frac{1}{2}(\omega^{-1})^{ij}\overline{\psi}^{j}\otimes \frac{\partial}{\partial x^{i}}.
\end{eqnarray}
Let us also define
\begin{equation}
I=\sum_{i}\mu_{i}^{2}.
\end{equation}
Then
\begin{eqnarray}
\lambda:=J(dI)=\mu_{i}\overline{\psi}^{i}.
\end{eqnarray}
Thus
\begin{equation}
\overline{\omega}=D\lambda.
\end{equation}

\subsection{An Equivariant 4-form in Local Riemannian Geometry}
Our next example is an equivariant 4-form defined on $M$. Consider the metric components $g_{ij}$ and the corresponding connection coefficients $\Gamma^{i}_{jk}$ in $U$. The following 4-form is globally defined on $M$
\begin{eqnarray}
\rho&:=&\frac{1}{2}\partial_{i}\partial_{j}g_{kl}\overline{\psi}^{i}\wedge\psi^{j}\wedge\overline{\psi}^{k}\wedge\psi^{l}
\end{eqnarray}
This 4-form is clearly closed and invariant under the action of $V$ since the coordinates $p^{i}$ appear only through their differentials $\overline{\psi}^{i}$.
Now we can form the equivariant extension of $\rho$ by determining
\begin{equation}
\overline{\rho}=\rho-ib^{i}\nu_{i}+f_{ij}b^{i}b^{j}
\end{equation}
subject to the conditions $\pounds_{i}\overline{\rho}=0$ and $D\overline{\rho}=0$. Here $\rho\in \Omega^{4}(U)$, $\nu^{i}\in \Omega^{2}(U)$ and $f_{ij}=f_{ji}\in \Omega^{0}(U)$.
Thus we have to solve the equations
\begin{eqnarray}
\pounds_{i}\nu_{j}&=&0 \label{e1}\\
\pounds_{k}f_{ij}&=&0\\
d\nu_{i}&=&-\iota_{i}\rho\label{e2}\\
df_{ij}&=&\frac{1}{2}\left\lbrace \iota_{i}\nu_{j}+\iota_{j}\nu_{i}\right\rbrace \label{e3}
\end{eqnarray}
Now notice that
\begin{eqnarray}
\iota_{n}\rho&=&\frac{1}{2}\lbrace
\partial_{n}\partial_{j}g_{kl}-\partial_{k}\partial_{j}g_{nl}\rbrace\psi^{j}\overline{\psi}^{k}\psi^{l}\nonumber\\
&=&\frac{1}{2}d\lbrace
\partial_{n}g_{kl}-\partial_{k}g_{nl}\rbrace\overline{\psi}^{k}\psi^{l}\nonumber\\
&=&\frac{1}{2}d\lbrace
\partial_{n}g_{kl}-\partial_{l}g_{kn}-\partial_{k}g_{nl}\rbrace\overline{\psi}^{k}\psi^{l}\nonumber\\
&=&-d\lbrace g_{ni}\Gamma^{i}_{kl}\overline{\psi}^{k}\psi^{l}\rbrace.
\end{eqnarray}
Choosing
\begin{eqnarray}
  \nu_{i} &=& g_{in}\Gamma^{n}_{kl}\overline{\psi}^{k}\psi^{l},
\end{eqnarray}
we have
\begin{eqnarray}
df_{ij}&=&\frac{1}{2}\left\lbrace \iota_{i}\nu_{j}+\iota_{j}\nu_{i}\right\rbrace \nonumber\\
&=&\frac{1}{2}\left\lbrace g_{in}\Gamma^{n}_{jl}+g_{jn}\Gamma^{n}_{il}\right\rbrace \psi^{l}\nonumber\\
&=& \frac{1}{2}dg_{ij}.
\end{eqnarray}
Hence we can take
\begin{equation}
f_{ij}=\frac{1}{2}g_{ij}.
\end{equation}
Thus we have the following equivariant extension of $\rho$
\begin{equation}
\overline{\rho}=\frac{1}{2}\partial_{i}\partial_{j}g_{kl}\overline{\psi}^{i}\psi^{j}\overline{\psi}^{k}\psi^{l}-
ib^{i}g_{in}\Gamma^{n}_{kl}\overline{\psi}^{k}\psi^{l}+\frac{1}{2}g_{ij}b^{i}b^{j}.
\end{equation}

\subsection{Integration}
Since our equivariant forms on $M$ depend on the coordinates $p^{i}$
only through $\overline{\psi}^{i}$ we have to regularize their
integrals. Thus we define
\begin{eqnarray}
\int \beta:=\int \left(\prod db^{i} \right)\int_{U\times
V}R(p)\beta.
\end{eqnarray}
Here $R(p)$ is a normalized Gaussian. Then it is not difficult to
see that the equivariant localization principle is valid for this
modified integral as well
\begin{equation}
\int \beta=\int \beta e^{tD\lambda},\;\;\;\;\mathrm{for}\;\;\;\;D\beta=0.
\end{equation}
In order to see that notice that
\begin{eqnarray}
  \int\, D\gamma &=& \int\, d\gamma \nonumber\\
   &=&  \int [db]\int_{M}R(p)b^{I}d[\gamma_{I,KL}\psi^{K}\overline{\psi}^{L}]\nonumber\\
  &=&\pm \int
  [db]b^{I}\int_{N}d[\gamma_{I,KL}\psi^{K}]\int_{V}R(p)\overline{\psi}^{L}=0.
\end{eqnarray}
Here in the last step we applied Stoke's theorem to the integral
over $N$ \footnote{If $N$ is not compact then we assume $\gamma$ to
have either compact support or to decay rapidly.}. Now applying this
result to the exact form $\beta (e^{tD\lambda}-1)$ we get the
desired result.

\subsection{Localization}

We can now localize integrals of the form
\begin{equation}
 \int \overline{\beta} e^{-\overline{\rho}}\,e^{\overline{\omega}}
\end{equation}
where $D\overline{\beta}=0$. Choosing the localizing factor as $e^{(s-1)D\lambda}$ we get
\begin{eqnarray}
 \int \overline{\beta} e^{-\overline{\rho}}\,e^{\overline{\omega}}&=&\int \overline{\beta} e^{-\overline{\rho}}\,e^{sD\lambda}\nonumber\\
&=&\int \overline{\beta} e^{-\overline{\rho}}\,e^{s\omega}e^{-isb^{i}\mu_{i}}\nonumber\\
&=&\int  \overline{\beta} e^{-\overline{\rho}}\left[1+s\omega+\ldots+\frac{s^{n}}{n!}\omega^{n} \right]e^{-isb^{i}\mu_{i}}\nonumber\\
&=&\int d^{n}b\, d^{n}x\, \sum_{k=1}^{n}\frac{s^{k}}{k!}[\overline{\beta} e^{-\overline{\rho}}\omega^{k}]_{top}\,e^{-isb^{i}\mu_{i}}
\end{eqnarray}
Here $[\overline{\alpha}]_{top}$ means the coefficient of the top
form in $\overline{\alpha}$. Under the non-degeneracy assumption
discussed in Sec.3.2 localization is onto $b=0$ and $\mu_{i}=0$.
Thus we get
\begin{eqnarray}
\sum_{k=1}^{n}\frac{s^{k}}{k!}\sum_{y_{c}}  [\overline{\beta}
e^{-\overline{\rho}}\omega^{k}]_{top}(y_{c})\left[
\frac{(2\pi)^{n}}{s^{n}\sqrt{|\det\,H(y_{c})|}}
e^{i\frac{\pi}{4}\xi}+O\left(\frac{1}{s^{n+1}}\right)\right].
\end{eqnarray}
Here $y_{c}$ stands for the critical points of $b^{i}\mu_{i}$ i.e.
$y_{c}=(b=0,x_{c})$ where $\mu(x_{c})=0$. $H$ is the Hessian of
$b^{i}\mu_{i}$ and $\xi$ is the signature of $H$ (number of positive
eigenvalues minus the number of negative ones). Notice that the
Hessian at $y_{c}$ is given by
\begin{eqnarray}
H(y_{c})=\left( \begin{array}{cc}
 0&\left. \frac{\partial \mu_{i}}{\partial x^{j}}\right\vert_{y_{c}}\\
\left. \frac{\partial \mu_{i}}{\partial x^{j}}\right\vert_{y_{c}} &0
 \end{array}\right)
\end{eqnarray}
and its determinant is $(\det\omega(y_{c}))^{2}$. So in the $s\rightarrow\infty$ limit the only contribution comes from the $k=n$ term
\begin{eqnarray}
\sum_{y_{c}} \frac{1}{n!}[\overline{\beta} e^{-\overline{\rho}}\omega^{n}]_{top}(y_{c})\, \frac{(2\pi)^{n}}{\det\,\omega(y_{c})}
e^{i\frac{\pi}{4}\xi}=\sum_{y_{c}} \frac{(2\pi)^{n}}{n!}  \overline{\beta}^{(0)}(y_{c})
e^{i\frac{\pi}{4}\xi}.
\end{eqnarray}
Here we used
\begin{equation}
[\overline{\beta}
e^{-\overline{\rho}}\omega^{n}]_{top}=\overline{\beta}^{(0)}\det\omega,
\end{equation}
where $\overline{\beta}^{(0)}$ is the zero form part of
$\overline{\beta}$, to cancel the Hessian in the denominator.

\subsection{Generalizations to Field Theory}

Let us now consider the set $\widetilde{U}$ which consists of all embeddings of an open set $\mathcal{O}\subset\mathbf{R}^{d+1}$ into the coordinate chart $U\subset N$. Let us also define $\widetilde{V}$ to be the set of all embeddings of $\mathcal{O}$ into $\mathbf{R}^{n}$. Consider the infinite dimensional space $\widetilde{M}:=\widetilde{U}\times \widetilde{V}$. The coordinates in $\widetilde{U}$ may be taken to be the functional values of the embedding function $\phi^{i}(t,x)$. Here, as usual in field theory, we treat both $i$ and $(t,x)$ as indices labeling the coordinates. We will denote the coordinates in $\widetilde{V}$ by $\pi^{i}(x,t)$. Differentials of these coordinates will be denoted by $\psi^{i}(t,x)$ and $\overline{\psi}^{i}(t,x)$, respectively. Clearly one can think of $\widetilde{V}$ as an abelian group acting on itself by translations. So we can define the action of $\widetilde{V}$ on $\widetilde{M}$ by fiber translations along the factor $\widetilde{V}$. All this is a more or less straightforward generalization of the finite dimensional case of the last section to the field theoretic framework. Therefore the action of the Cartan differential on the generators of the Cartan complex is defined by
\begin{eqnarray}
D\phi^{i}(t,x)&=&\psi^{i}(t,x)\\
D\psi^{i}(t,x)&=&0\\
D\overline{\psi}^{i}(t,x)&=&-ib^{i}(t,x)\\
Db^{i}(t,x)&=&0.
\end{eqnarray}
But these are nothing but the Parisi-Sourlas SUSY transformations of stochastic quantization. 

The FP action for a non-linear sigma model is given by
\begin{equation}
S_{FP}=\overline{\rho}+\overline{\omega}
\end{equation}
where
\begin{equation}
\overline{\rho}=\int d^{d}x dt \,\left[ \frac{1}{2}\partial_{i}\partial_{j}g_{kl}(\phi)\overline{\psi}^{i}\psi^{j}\overline{\psi}^{k}\psi^{l}-
ib^{i}g_{kn}(\phi)\Gamma^{n}_{il}(\phi)\overline{\psi}^{k}\psi^{l}+\frac{1}{2}g_{ij}b^{i}b^{j}\right],
\end{equation}
and
\begin{eqnarray}
\overline{\omega}=\omega-i\int d^{d}x dt\, b^{i}\mu_{i}.
\end{eqnarray}
Here
\begin{eqnarray}
\omega=-\int d^{d}x dt\,
\overline{\psi}^{i}\left[g_{ij}(\phi)\frac{d}{dt}+\partial_{j}g_{ik}(\phi)\frac{d\phi^{k}}{dt}+
 \frac{\delta^{2}
S}{\delta\phi^{i}(t,x)\delta\phi^{j}(t,x)}\right]\psi^{j}
\end{eqnarray}
and
\begin{eqnarray}
\mu_{i}[\phi;x,t]=g_{ij}\frac{d\phi^{j}}{dt}+ \frac{\delta
S}{\delta\phi^{i}(t,x)},
\end{eqnarray}
where $\partial_{i}=\frac{\partial}{\partial \phi^{i}(t,x)}$. It is
easy to see that $\overline{\omega}$ is exact
\begin{equation}
\overline{\omega}=D\lambda,\;\;\;\;\lambda=\int d^{d}x dt\,
\overline{\psi}^{i}\left[g_{ij}\frac{d\phi^{j}}{dt}+ \frac{\delta
S}{\delta\phi^{i}(t,x)} \right].
\end{equation}
Thus we see that $S_{FP}$ is the sum of an equivariant
(pre)symplectic form and an equivariant 4-form. So the Boltzmann
factor of our model is a closed equivariant form as well. This
characterization of $S_{FP}$ allows us to use equivariant
localization principle to localize path integrals with closed
insertions as we did in the last section for the finite dimensional
case. However in stochastic quantization our real interest is in
(the large $t$ limit of) the path integrals of the form
\begin{eqnarray}
<\phi^{i_{1}}(t,x_{1})\ldots
\phi^{i_{n}}(t,x_{n})>=\frac{1}{Z_{FP}}\int
\phi^{i_{1}}(t,x_{1})\ldots
\phi^{i_{n}}(t,x_{n})e^{-\overline{\rho}-\overline{\omega}}.
\end{eqnarray}
Clearly the insertion
\begin{equation}
 \phi^{i_{1}}(t,x_{1})\ldots \phi^{i_{n}}(t,x_{n})
\end{equation}
is not a closed equivariant form. So a straightforward application of equivariant localization
does not seem possible in the case of physical importance. However we will show that if one assumes
the dimension of the target space to be arbitrarily large and considers only the insertions which satisfy
some kind of large $N$ factorization property then it is possible to apply the localization principle.
In what follows we will carry this program for matrix models.

\section{Application to Matrix Models}

Having established the equivariant cohomological character of the
Parisi-Sourlas supersymmetry of stochastic quantization we can apply
localization methods to the FP field theory. Clearly one can use
equivariant localization principle to calculate the partition
function of the FP theory. However, one has to be careful at this
point since the equivariant 2-form appearing in the action is only a
pre-symplectic form, i.e. $\omega$ is degenerate at certain field
configurations. In stochastic quantization our real interest is in
the correlations rather than the partition function. The problem
encountered in applying equivariant localization principle to the
correlations is that the latter are not in general equivariantly
closed forms. We will show that for matrix models we can remedy the
situation by going to the large $N$ limit. From this point on we
will consider only the case of Hermitian matrix models for which the
target space is flat.

\subsection{Localization of $Z_{FP}$}

Consider a $U(N)$-invariant Hermitian matrix model in $d$ Euclidean
dimensions. For definiteness let us take the action of the model to
be
\begin{equation}
\int d^{d}x\, \mathrm{tr} \left[\frac{1}{2}\partial_{\mu}\phi\partial_{\mu}\phi+V(\phi) \right],
\end{equation}
where
\begin{equation}
V(\phi)=\sum_{k\geq 2}\frac{g_{k}}{k}\phi^{k}.
\end{equation}
The Langevin equation for this model is
\begin{eqnarray}
\eta^{a}_{b}(t,x)&=&\partial_{t}\phi^{a}_{b}(t,x)+\frac{\delta S}{\delta \phi^{b}_{a}(t,x)}\\
&=&\partial_{t}\phi^{a}_{b}(t,x)-\partial_{x}^{2}\phi^{a}_{b}(t,x)+\sum_{k}g_{k}[ \phi^{k-1}(t,x)]^{a}_{b}.
\end{eqnarray}
The corresponding Fokker-Planck action is derived in Appendix B. The result is
\begin{eqnarray}
S_{FP}=\int dt d^{d}x \tr \left[  \frac{b^{2}}{2}
+ib\eta_{\phi}+\overline{\psi}\left(\partial_{t}-\partial^{2}_{x}
\right)\psi+\sum_{k}g_{k}\sum_{m+n=k-2}\overline{\psi}\phi^{m}\psi\phi^{n}\right] \nonumber\\
\end{eqnarray}
where
\begin{equation}
(\eta_{\phi})^{a}_{b}=
\partial_{t}\phi^{a}_{b}-\partial_{x}^{2}\phi^{a}_{b}+\sum_{k}g_{k} (\phi^{k-1})^{a}_{b}
\end{equation}
This action is invariant under the Parisi-Sourlas SUSY transformation
\begin{eqnarray}
D\phi^{a}_{b}&=&\psi^{a}_{b}\\
D\psi^{a}_{b}&=&0\\
D\overline{\psi}^{a}_{b}&=&-ib^{a}_{b}\\
Db^{a}_{b}&=&0.
\end{eqnarray}
In fact,
\begin{eqnarray}
S_{FP}=\int dt d^{d}x\,
D\mathrm{tr}(-\eta\overline{\psi}+ib\overline{\psi}).
\end{eqnarray}
A straightforward generalization of the results of the last section allows us to identify the SUSY operator as a Cartan derivative. Moreover
\begin{eqnarray}
\int dt d^{d}x\,\tr  \left[ ib\eta_{\phi} +\overline{\psi}\left(\partial_{t}-\partial^{2}_{x}
\right)\psi +\sum_{k}g_{k}\sum_{m+n=k-2}\overline{\psi}\phi^{m}\psi\phi^{n}\right]
\end{eqnarray}
is annihilated by $D$ and therefore it can be identified as a closed equivariant 2-form. Thus we have a (pre)symplectic form
\begin{equation}
\omega=\int dt d^{d}x\,\tr\left[
\overline{\psi}\left(\partial_{t}-\partial^{2}_{x} \right)\psi
+\sum_{k}g_{k}\sum_{m+n=k-2}\overline{\psi}\phi^{m}\psi\phi^{n}\right]
\end{equation}
and the moment map $\mu$ for fiber translations is given by
\begin{equation}
\mu=-\eta_{\phi}.
\end{equation}
So we can write $S_{FP}$ as
\begin{eqnarray}
 S_{FP}=\frac{1}{2}b\cdot b+\omega-ib\cdot \mu=\frac{1}{2}b\cdot b+\overline{\omega}
\end{eqnarray}
where we used the shorthand
\begin{equation}
 F\cdot G=\int dt\,d^{d}x\,\tr\left[  F(t,x)G(t,x)\right] .
\end{equation}

Now we can apply the equivariant localization principle to the partition function
\begin{eqnarray}
Z_{FP}=\int \mathcal{D}b\, e^{-N\,b\cdot b}\int
e^{-N\overline{\omega}}=\lim_{s\rightarrow\infty}\int \mathcal{D}b\, e^{-N\,b\cdot b}\int
e^{-N\overline{\omega}}e^{-N(s-1)D\lambda}.
\end{eqnarray}
Here the fields and the coupling constants are appropriately scaled in order to get an overall factor of $N$ in front of the action.
Here the degeneracy of the pre-symplectic form $\omega$ is not harmful since the integrand is basically proportional to the determinant of $\omega$ which vanishes at the points where the latter is degenerate.

We can localize $Z_{FP}$ either by proceeding as we did in Sec.4.4 or by observing that in the integral for $Z_{FP}$ the convergence factor $e^{-N\,b\cdot b}$ can be omitted without changing the value of the integral. The derivation of this result is given in Appendix C and relies heavily on the localization of the integral. After this modification the integral over $b$ gives a delta function
\begin{eqnarray}\label{count1}
Z_{FP}=\int e^{-sN\omega}\delta(sN\mu^{a}_{b}).
\end{eqnarray}
But the top form in the differential form $e^{-Ns\omega}$ is
\begin{equation}
\det\left[sN\frac{\delta \mu}{\delta \phi} \right].
\end{equation}
So we get
\begin{equation}\label{count2}
Z_{FP}=\int \mathcal{D}\phi \,\sum_{\overline{\phi}}\delta(\phi^{a}_{b}(t,x)-\overline{\phi}^{a}_{b}(t,x))=\sum_{\overline{\phi}}1
\end{equation}
where $\overline{\phi}$ is the solution of $\mu=0$, i.e.
\begin{eqnarray}
\partial_{t}\overline{\phi}^{a}_{b}(t,x)-\partial_{x}^{2}\overline{\phi}^{a}_{b}(t,x)+\sum_{k}g_{k}(\overline{\phi}^{k-1}(t,x))^{a}_{b}=0.
\end{eqnarray}
Thus we can interpret $Z_{FP}$ as the volume of the space of solutions of $\mu=0$. The solution to this equation can be written as
\begin{equation}
\overline{\phi}(x,t)=e^{iP\cdot x}A(P;t)e^{-iP\cdot x}
\end{equation}
with
\begin{eqnarray}
 (P^{\mu})^{a}_{b}=p_{\mu}^{a}\delta^{a}_{b}
\end{eqnarray}
and
\begin{equation}\label{m1}
\partial_{t}A^{a}_{b}(P;t)+A^{a}_{b}(P;t)(p^{a}-p^{b})^{2}+\sum_{k}g_{k}(A^{k-1}(P;t))^{a}_{b}=0.
\end{equation}
Notice that these solutions are parametrized by the diagonal matrices $P_{\mu}$ and the matrix $A(P,t)$. 
We will assume that in the large $N$ limit these are the only solutions of $\mu=0$. The justification of this assumption, which may be based on the methods developed in \cite{jl, bardakci, halpern1}, will be given elsewhere.
In the following we will take the measure on $\mu^{-1}(0)$ in the form
\begin{eqnarray}
Z_{FP}=\int d\mu(P) \int d\mu_{P}(A)
\end{eqnarray}
where $d\mu(P)$ is a suitable measure on the space of diagonal matrices $P_{\mu}$
and $d\mu_{P}(A)$ is the measure on the space of solutions of (\ref{m1}) for fixed $P_{\mu}$. We will discuss $d\mu(P)$ in the next section. For now let us concentrate on $d\mu_{P}(A)$. Let us define
\begin{eqnarray}
\widetilde{\mu}^{a}_{b}[P,A]&=&\partial_{t}A^{a}_{b}(t)+(p^{a}-p^{b})^{2}A^{a}_{b}(t)+\sum_{k}g_{k}(A^{k-1}(t))^{a}_{b}\label{red1}\label{sim}
\end{eqnarray}
where there is no sum over repeated indices. Notice that $\widetilde{\mu}$ has a dependence on $P$.

In analogy with (\ref{count1}) and (\ref{count2}) the following measure counts the number of solutions of $\widetilde{\mu}^{a}_{b}[P,A]=0$
\begin{equation}
  \mathcal{D}A\,\mathcal{D}c\,\mathcal{D}\overline{c}\;\delta(sNV\widetilde{\mu}[P,A])e^{-sNV\widetilde{\omega}},
\end{equation}
where
\begin{eqnarray}
\widetilde{\omega}=NV\int dt\,\tr\left[  \overline{c}\,\partial_{t}c+\sum_{k}g_{k}\sum_{m+n=k-2}\overline{c}A^{m}\,c\,A^{n} \right]-\sum_{a,b}\overline{c}^{a}_{b} (p^{a}-p^{b})^{2}c^{b}_{a}.\nonumber\\\label{red2}\label{mom}
\end{eqnarray}
Thus we can write the partition function as
\begin{eqnarray}
Z_{FP}=\int d\mu(P) \int \mathcal{D}A\,\mathcal{D}c\,\mathcal{D}\overline{c}\,\delta(sNV\widetilde{\mu})e^{-sNV\widetilde{\omega}}.
\end{eqnarray}

\subsection{Localization of $\widetilde{Z}_{FP}$}

In this section we will show that the expression for $Z_{FP}$ that we derived in the last section is nothing but the quenched average of the partition function of the FP action of the Eguchi-Kawai model corresponding to the original matrix model. So again let us start with a matrix model whose action is
\begin{equation}
\int d^{d}x\, \mathrm{tr} \left[\frac{1}{2}\partial_{\mu}\phi\partial_{\mu}\phi+V(\phi) \right],
\end{equation}
The corresponding Eguchi-Kawai model is a zero dimensional matrix model with the action
\begin{eqnarray}
S_{EK}&=&\sum_{a,b}\frac{1}{2}(p_{a}-p_{b})^{2}A^{a}_{b}A^{b}_{a}+\sum_{k}\frac{g_{k}}{k}\tr\,A^{k}\\
&=&\tr \left( -\frac{1}{2} \left[P_{\mu},A \right]\left[P_{\mu},A \right]+\sum_{k}\frac{g_{k}}{k}\,A^{k}\right).
\end{eqnarray}
Here $(P_{\mu})^{a}_{b}=p_{\mu}^{a}\delta^{a}_{b}$.
The FP action corresponding to this model is given by (see Appendix B)
\begin{eqnarray}
(S_{EK})_{FP}=\int dt\,
\tr\,\left[ \frac{1}{2}\sigma^{2}-i\tr\,(\sigma\widetilde{\mu})\right]+\widetilde{\omega} .
\end{eqnarray}
where $\widetilde{\mu}$ and $\widetilde{\omega}$ are given by
(\ref{red1}) and (\ref{red2}), respectively. The action of the
Cartan derivative on the fields is
\begin{eqnarray}
  DA(t) &=& c(t),\,\,\;\;\; Dc(t)=0 \\
  D\overline{c}(t) &=&-i\sigma(t), \,\,\;\;\;
  D\sigma(t)=0.
\end{eqnarray}
Applying the equivariant localization principle to the partition
function of this action and eliminating the quadratic term in $\sigma$ in exactly the same way as in the previous section we get
\begin{eqnarray}
\widetilde{Z}_{FP}=\int \mathcal{D}A\mathcal{D}c\mathcal{D}\overline{c}\,\delta(sN\widetilde{\mu})e^{sN\widetilde{\omega}}
\end{eqnarray}
Thus one has
\begin{equation}
Z_{FP}=\int d\mu(P) \widetilde{Z}_{FP}.
\end{equation}
Now we will choose the measure $d\mu(P)$ in such a way that $Z_{FP}=1$. Let
\begin{equation}
 dP=\prod_{\mu,a}dp_{\mu}^{a}
\end{equation}
and let $\rho(P)$ be a Gaussian normalized with respect to $dP$. Now choosing
\begin{equation}
 d\mu(P)=\frac{1}{\widetilde{Z}_{FP}}\rho(P)dP
\end{equation}
we get $Z_{FP}=1$.

\subsection{Correlations of $S_{FP}$}

In stochastic quantization our real interest is in the $U(N)$ invariant correlations of the model. The
obstruction for the use of equivariant localization principle in
this case is that an insertion of the form
\begin{eqnarray}
F(t;x_{1},\ldots,x_{n})=\frac{1}{N}\tr\left[\phi(t,x_{1})\ldots \phi(t,x_{n}) \right]
\end{eqnarray}
is not a closed equivariant form. Therefore one cannot directly apply the localization principle to the path integral
\begin{eqnarray}\label{w1}
<F(t;x_{1},\ldots,x_{n})>=\frac{1}{Z}\int \frac{1}{N}\tr\left[\phi(t,x_{1})\ldots \phi(t,x_{n}) \right]\, e^{-N\,b\cdot b} e^{N\overline{\omega}}\nonumber\\
\end{eqnarray}
to get a formula like
\begin{eqnarray}\label{w2}
<F(t;x_{1},\ldots,x_{n})>=\frac{1}{Z}\int \frac{1}{N}\tr\left[\phi(t,x_{1})\ldots \phi(t,x_{n}) \right]e^{-Ns\omega}\delta(sN\mu^{a}_{b}).\nonumber\\
\end{eqnarray}
However, we will show in this section that the passage from (\ref{w1}) to (\ref{w2}) is in fact possible in the large $N$ limit.

First of all notice that $F$, $DF$, $\frac{\lambda}{N}$ and $\frac{D\lambda}{N}$ are all $U(N)$ invariant observables and their normalizations ensure that they satisfy the large $N$ factorization. Consider
\begin{eqnarray}
&&\frac{1}{Z_{FP}}\int F e^{-N\,b\cdot b} e^{N\overline{\omega}}\left[e^{\frac{s}{N}D\lambda}-1 \right]=\nonumber\\
&& =\frac{1}{Z_{FP}}\int F e^{-N\,b\cdot b} e^{N\overline{\omega}}\left[\frac{s}{N}D\lambda+\frac{1}{2}\left(\frac{s}{N} \right)^{2}D\lambda D\lambda+\ldots  \right]\nonumber\\
&&=\frac{1}{Z_{FP}}\int F e^{-N\,b\cdot b} e^{N\overline{\omega}}D\left[\frac{s}{N}\lambda+\frac{1}{2}\left(\frac{s}{N} \right)^{2}\lambda D\lambda+\ldots  \right]\nonumber\\
&&=-\frac{1}{Z_{FP}}\int (DF) e^{-N\,b\cdot b} e^{N\overline{\omega}}\left[\frac{s}{N}\lambda+\frac{1}{2}\left(\frac{s}{N} \right)^{2}\lambda D\lambda+\ldots  \right]\nonumber\\
&&=\sum_{k=1}^{\infty}\frac{s^{k}}{k!}\left\langle(DF)\left(
\frac{\lambda}{N}\right)
\left(\frac{D\lambda}{N}\right)^{k-1}\right\rangle
\end{eqnarray}
Using the large $N$ factorization we see that each term in the sum contains factors of $<DF>$ and $<\frac{\lambda}{N}>$. But notice that
\begin{equation}
\frac{1}{N}\lambda=\int\,dt d^{d}x \,\frac{1}{N}\tr\left[ \mu(t,x)\overline{\psi}(t,x)\right]
\end{equation}
and
\begin{equation}
DF=\sum_{l=1}^{n}\frac{1}{N}\tr\left[\phi(t,x_{1})\ldots\psi(t,x_{l})\ldots\phi(t,x_{n}) \right]
\end{equation}
So both expectation values are, at least perturbatively, zero. Thus we establish the validity of the equivariant localization principle in the large $N$ limit
\begin{eqnarray}
\int F e^{-N\,b\cdot b} e^{-N\overline{\omega}}=\int F e^{-N\,b\cdot b} e^{-N\overline{\omega}}e^{-\frac{s-1}{N}D\lambda}
\end{eqnarray}
Now, we can repeat the argument of the last subsection and eliminate the convergence factor for the $b$ integral. Then integration over the $b$ field gives us (\ref{w2}) which is localized on the zeroes of $\mu^{a}_{b}$. But recall that these zeroes are given by
\begin{equation}
\overline{\phi}^{a}_{b}(t,x)=M^{a}_{b}(P;t)e^{i(p^{a}-p^{b})\cdot x}
\end{equation}
with
\begin{equation}
\partial_{t}M^{a}_{b}(P;t)+M^{a}_{b}(P;t)(p^{a}-p^{b})^{2}+\sum_{k}g_{k}(M^{k-1}(P;t))^{a}_{b}=0.
\end{equation}
So we have
\begin{eqnarray}
<F>&=&\frac{1}{Z_{FP}}\sum_{\overline{\phi}}\sum_{a_{1},\ldots,a_{n}}\frac{1}{N}\,M^{a_{1}}_{a_{2}}(P;t)M^{a_{2}}_{a_{3}}(P;t)\ldots M^{a_{n}}_{a_{1}}(P;t)\times\nonumber\\
&&\times e^{i(p^{a_{1}}-p^{a_{2}})\cdot x_{1}}\,
e^{i(p^{a_{2}}-p^{a_{3}})\cdot x_{2}}\ldots e^{i(p^{a_{n}}-p^{a_{1}})\cdot x_{n}}.
\end{eqnarray}
But this sum can also be written as
\begin{eqnarray}
&&\frac{1}{Z_{FP}}\int d\mu(P)\sum_{a_{1},\ldots,a_{n}} e^{i(p^{a_{1}}-p^{a_{2}})\cdot x_{1}}\,
e^{i(p^{a_{2}}-p^{a_{3}})\cdot x_{2}}\ldots e^{i(p^{a_{n}}-p^{a_{1}})\cdot x_{n}}\times\nonumber\\
&& \times\int \mathcal{D}A\; \sum_{M(P,t)}\delta(A(t)-M(P;t))\frac{1}{N}\,A^{a_{1}}_{a_{2}}(t)A^{a_{2}}_{a_{3}}(t)\ldots A^{a_{n}}_{a_{1}}(t)
\end{eqnarray}
Using the standard delta function identities the $\mathcal{D}A$ integral can be written as
\begin{eqnarray}
\int \mathcal{D}A\,\mathcal{D}c\,\mathcal{D}\overline{c}\,\frac{1}{N}\,\left[ A^{a_{1}}_{a_{2}}(t)A^{a_{2}}_{a_{3}}(t) \ldots A^{a_{n}}_{a_{1}}(t)\right]  \delta(sNV\widetilde{\mu})e^{-sNV\widetilde{\omega}}
\end{eqnarray}
where
\begin{eqnarray}
\widetilde{\mu}^{a}_{b}[P,A]&=&-\left[ \partial_{t}A^{a}_{b}(t)+(p^{a}-p^{b})^{2}A^{a}_{b}(t)+\sum_{k}g_{k}(A^{k-1}(t))^{a}_{b}\right] \\
\widetilde{\omega}&=&\int dt\,\tr\left[  \overline{c}\,\partial_{t}c+\sum_{k}g_{k}\sum_{m+n=k-2}\overline{c}A^{m}\,c\,A^{n} \right]-\sum_{a,b}\overline{c}^{a}_{b} (p^{a}-p^{b})^{2}c^{b}_{a}.\nonumber\\
\end{eqnarray}
and there is no summation over repeated indices.

So we get
\begin{eqnarray}\label{l1}
<F>=\frac{1}{Z_{FP}}\int d\mu(P)\int \mathcal{D}A\,\mathcal{D}c\,\mathcal{D}\overline{c}\,\widetilde{F}\,\delta(sN\widetilde{\mu})e^{-sN\widetilde{\omega}}
\end{eqnarray}
where
\begin{eqnarray}
\widetilde{F}=\frac{1}{N}\,\tr\left(\widetilde{\phi}(t,x_{1})\ldots \widetilde{\phi}(t,x_{n})  \right),
\end{eqnarray}
and
\begin{equation}
\widetilde{\phi}(t,x)=e^{iP\cdot x}A(t)e^{-iP\cdot x},\;\;\;(P^{\mu})^{a}_{b}=p_{\mu}^{a}\delta^{a}_{b}.
\end{equation}

\subsection{Correlations of $(S_{EK})_{FP}$}
Now let us consider correlations of $(S_{EK})_{FP}$ with insertions of the form
\begin{eqnarray}
\widetilde{F}(t,x_{1},\ldots,x_{n})=\frac{1}{N}\tr\left[\widetilde{\phi}(t, x_{1})\ldots \widetilde{\phi}(t, x_{n}) \right]
\end{eqnarray}
where the quenched field $\widetilde{\phi}(x)$ is defined as
\begin{eqnarray}
\widetilde{\phi}(x)=e^{iP\cdot x}A(t)e^{-iP\cdot x},\;\;\;\;\;  (P^{\mu})^{a}_{b}=p_{\mu}^{a}\delta^{a}_{b}.
\end{eqnarray}
Here, again the obstruction for the use of equivariant localization
is the non-closedness of the insertion. In order to eliminate this
obstruction we will assume that the integrals of the form
\begin{eqnarray}\label{m2}
\frac{1}{Z}\int d\mu(P)\int \mathcal{D}A\,\mathcal{D}c\,\mathcal{D}\overline{c}\,\widetilde{F}\,e^{-(S_{EK})_{FP}}
\end{eqnarray}
\begin{equation}
 Z=\int d\mu(P)\int \mathcal{D}A\,\mathcal{D}c\,\mathcal{D}\overline{c}\,e^{-(S_{EK})_{FP}}=\int e^{-S_{FP}}
\end{equation}
obey the large $N$ factorization \cite{gk}. Here the matrices $P$ are treated as usual degrees of freedom; they are not quenched yet. Also notice that the theory defined by (\ref{m2}) is invariant under conjugation of $P$ and $A$ by $[U(1)]^{N}$ matrices. Under the factorization assumption one can use the same argument that was used above in the case of $S_{FP}$ to show that the equivariant localization is valid for (\ref{m2}). In particular one can perform the $\sigma$ integral without the convergence factor and get the result
\begin{eqnarray}
\frac{1}{Z_{FP}}\int d\mu(P)\int \mathcal{D}A\,\mathcal{D}c\,\mathcal{D}\overline{c}\,\widetilde{F}\,\delta(sNV\widetilde{\mu})e^{-sNV\widetilde{\omega}}.
\end{eqnarray}
Comparing this with (\ref{l1}), the localized expectation of $F$, we get
\begin{eqnarray}
 <F>_{S_{FP}}=\frac{1}{Z}\int d\mu(P)\int \mathcal{D}A\,\mathcal{D}c\,\mathcal{D}\overline{c}\,\widetilde{F}\,e^{-(S_{EK})_{FP}}.
\end{eqnarray}
Now using the explicit form of $d\mu(P)$ we obtain
\begin{equation}
<F>_{S_{FP}}=\int dP\,\rho(P)<\widetilde{F}>_{(S_{EK})_{FP}}.
\end{equation}
This is nothing but the quenching prescription written in terms of FP actions. Taking the large $t$ limit we get the desired result
\begin{equation}
 <F>=\int dP\,\rho(P) <\widetilde{F}>_{S_{EK}}.
\end{equation}

\section {Conclusions}

In conclusion, we showed that the Parisi-Sourlas supersymmetry of FP actions that arise in stochastic quantization of linear and non-linear sigma models can be interpreted as equivariant cohomology based on the action of a translation group on an extended configuration space. We applied equivariant localization principle to localize closed observables of FP actions. However we also noticed that the color invariant observables are not closed and that this fact obstructs the use of localization principle in the situation of physical interest. However we showed that this obstruction can be eliminated in the large $N$ limit of matrix models by using the factorization property, and consequently we derived the quenched Eguchi-Kawai models by equivariant localization.

\paragraph{Acknowledgment:} The author would like to thank T. Dereli for useful comments.

\appendix

\section{Derivation of $S_{FP}$}

Here we will follow \cite{gozzi} with minor modifications. Let $Z[J]$ be the generating function for the correlations of the
Langevin equation
\begin{equation}
\frac{\partial\phi(t,x)}{\partial t}+\frac{\delta S}{\delta
\phi(t,x)}=\eta(t,x).
\end{equation}
Denoting the solution of this equation by $\phi_{\eta}$ we get
\begin{equation}
Z[J]=\int \mathcal{D}\phi\mathcal{D}\eta\,\delta(\phi-\phi_{\eta})\exp\left[ {-\int dt d^{d}x\, \frac{1}{2}\eta^{2}-J\phi}\right] .
\end{equation}
Using the identity
\begin{equation}
\delta(\phi-\phi_{\eta})=\delta(\eta-\eta_{\phi})\det\left[ \frac{\delta \eta_{\phi}}{\delta \phi}\right],
\end{equation}
where $\eta_{\phi}$ is defined by the Langevin equation, and integrating over $\eta$ we get
\begin{equation}
Z[J]=\int \mathcal{D}\phi\,\det\left[ \frac{\delta \eta_{\phi}}{\delta \phi} \right] e^{-\int dt d^{d}x\, \frac{1}{2}\eta^{2}-J\phi}.
\end{equation}
Using the path integral representation of the functional determinant
\begin{equation}
\det\left[  \frac{\delta \eta}{\delta \phi}\right]=\int [d\psi][d\overline{\psi}]\exp\left[- \int dt d^{d}xdt' d^{d}x'\; \overline{\psi}(x,t)\left[  \frac{\delta \eta(x',t')}{\delta \phi(t,x)}\right]\psi(x',t')\right]  .
\end{equation}
and introducing the auxiliary field $b$ we get
\begin{equation}
Z[J]=\int
\mathcal{D}\phi\mathcal{D}b\mathcal{D}\psi\mathcal{D}\overline{\psi}\,
e^{-\int dt d^{d}x\,
(\frac{1}{2}b^{2}+ib\eta_{\phi}+\overline{\psi}\frac{\delta
\eta_{\phi}}{\delta \phi}\psi -J\phi)}.
\end{equation}
Using
\begin{equation}
\frac{\delta \eta(t',x')}{\delta \phi(t,x)}=\frac{\partial}{\partial
t}\delta(t-t')\delta(x-x')+\frac{\delta^{2}S}{\delta\phi(t',x')\delta\phi(t,x)}.
\end{equation}
we can read off $S_{FP}$ as
\begin{eqnarray}
S_{FP}&=&\int d\tau d^{d}x\, \frac{1}{2}b^{2}(t,x)+ib(t,x)\left[
\frac{\partial\phi(t,x)}{\partial t}+\frac{\delta
S[\phi]}{\delta\phi(t,x)}\right]\\ \nonumber
&&+\overline{\psi}(t,x)\left[\partial_{t}\delta(t-t')\delta^{d}(x-x')+\frac{\delta
S[\phi]}{\delta\phi(t',x')\delta\phi(t,x)}\right]\psi(t',x')
\end{eqnarray}

\section{FP Actions for Matrix Models}

The derivation given above can easily be generalized to
multi-component fields. In particular consider a matrix model with the
action 
\begin{equation}
 S=\int d^{d}x \tr\left[\frac{1}{2}\partial_{\mu}\phi \partial^{\mu}\phi+V(\phi) \right]
\end{equation}
where
\begin{equation}
 V(\phi)=\sum_{k\geq 2}\frac{g_{k}}{k}\phi^{k}.
\end{equation}
The Langevin equation is given by
\begin{equation}
 \partial_{t}\phi^{a}_{b}(t,x)+\frac{\delta S}{\delta \phi^{b}_{a}(t,x)}=\eta^{a}_{b}(t,x).
\end{equation}
and the corresponding FP action is ($z=(t,x)$)
\begin{eqnarray}
&&S_{FP}=\int dz\,
\frac{1}{2}b^{b}_{a}(z)b^{a}_{b}(z)+ib^{b}_{a}(z)\left[
\frac{\partial\phi^{a}_{b}}{\partial t}+\frac{\delta
S[\phi]}{\delta\phi^{b}_{a}(z)}\right]+\nonumber\\
&&+\int dz dz' \overline{\psi}^{b}_{a}(z)\left[\partial_{t}\delta(z-z')\delta^{a}_{d}\delta^{c}_{b}+
\frac{\delta S[\phi]}{\delta\phi^{d}_{c}(z')\delta\phi^{b}_{a}(z)}\right]\psi^{d}_{c}(z')\nonumber\\
\end{eqnarray}
Notice that
\begin{equation}
\frac{\delta S}{\delta \phi^{b}_{a}(z)}=-\partial^{2}_{x}\phi^{a}_{b}(z)+\sum_{k\geq 2} g_{k}(\phi^{k-1}(z))^{a}_{b}
\end{equation}
and
\begin{eqnarray}
\frac{\delta^{2} S}{\delta \phi^{b}_{a}(z)\delta \phi^{d}_{c}(z')}=\left[ -\partial^{2}_{x}\delta^{c}_{b}\delta^{a}_{d}+\sum_{k\geq 2}g_{k}
\sum_{m+n=k-2}(\phi^{m})^{a}_{d}(\phi^{n})^{c}_{b}\right] \delta(z-z')\nonumber\\
\end{eqnarray}
So we have
\begin{eqnarray}
S_{FP}=\int dz\, \tr\left[\frac{1}{2}b^{2}+ib\eta+\overline{\psi}(\partial_{t}-\partial^{2}_{x})\psi+\sum_{k\geq2}g_{k}\sum_{m+n=k-2}\overline{\psi}\phi^{m}\psi\phi^{n} \right].\nonumber\\
\end{eqnarray}
By similar calculations the FP action for the Eguchi-Kawai model
\begin{equation}
 S_{EK}=\tr \left( -\frac{1}{2} \left[P_{\mu},A \right]\left[P_{\mu},A \right]+\sum_{k}\frac{g_{k}}{k}\,A^{k}\right)
\end{equation}
is given by
\begin{eqnarray}
(S_{EK})_{FP}&=&\int dt \,\tr\left[\frac{1}{2}\sigma^{2}+i\sigma \widetilde{\eta}+  \overline{c}\,\partial_{t}c+\sum_{k}g_{k}\sum_{m+n=k-2}\overline{c}A^{m}\,c\,A^{n} \right.\nonumber\\
&&\left.-\sum_{a,b}\overline{c}^{a}_{b} (p^{a}-p^{b})^{2}c^{b}_{a} \right]
\end{eqnarray}
where
\begin{equation}
\widetilde{\eta}^{a}_{b}=\partial_{t}A^{a}_{b}(t)+(p^{a}-p^{b})^{2}A^{a}_{b}(t)+\sum_{k}g_{k}(A^{k-1}(t))^{a}_{b}
\end{equation}

\section{Elimination of the Convergence Factor}
Consider the integral \cite{akant}
\begin{equation}\label{inte}
 \int \alpha e^{-\overline{\rho}-\overline{\omega}+sD\lambda}
\end{equation}
with $D\alpha=0$ and $\overline{\rho}=D\beta$. Notice that in the case of matrix models $\overline{\rho}$ is indeed exact
\begin{equation}
\overline{\rho}=-iD\int d\tau\,d^{d}x\;\tr\left[b(\tau,x)\overline{\psi}(\tau,x)\right]
\end{equation}
Let $B'$ be an equivariant neighborhood of $\mu^{-1}(0)$ and let
$B\supset B'\supset \mu^{-1}(0)$. Consider a bump function $u$ which
is $1$ on $B'$ and vanishes outside $B$. Because of localization we
can insert a factor of $u$ in the integrand
\begin{eqnarray}
\Delta= \int_{M} \alpha (e^{-D\beta}-1)e^{-\overline{\omega}} e^{sD\lambda}= \int_{M} \alpha (e^{-D\beta}-1)e^{-\overline{\omega}} e^{sD\lambda}u
\end{eqnarray}
Moreover we can restrict the integral to $B$. Thus
\begin{eqnarray}
&&\int_{B} \left( e^{D\beta}-1\right) e^{-\overline{\omega}+sD\lambda}=\nonumber\\
&&\int_{B} D\left(\beta+\frac{1}{2}\beta D\beta+\ldots \right) e^{-\overline{\omega}+sD\lambda}u=\nonumber\\
&&\int_{B} d\left[ \left(\beta+\frac{1}{2}\beta D\beta+\ldots\right)e^{-\overline{\omega}+sD\lambda}u\right]+\nonumber\\
&&+\int_{B}
\left(\beta+\frac{1}{2}\beta D\beta+\ldots\right)
e^{-\overline{\omega}+sD\lambda}(du)
\end{eqnarray}
The first term is a surface term on $B$ where $u$ vanishes. On the other hand in the large $s$
limit the second term can be restricted onto $B'\times V$
where $du=0$. Thus we conclude that the convergence factor may be
omitted in (\ref{inte}). Notice that in this derivation the exactness of $\overline{\rho}$ is essential. Otherwise one would need the machinery developed in \cite{witten1, KJ}.

\end{document}